\documentstyle[aps,multicol,prl,epsf]{revtex}
\begin{document}
\draft
\title{Quantized pumped charge due to surface acoustic waves in a 1D channel }
\author{Amnon Aharony and O. Entin-Wohlman}
\address{School of Physics and Astronomy, Raymond and Beverly Sackler
Faculty of Exact Sciences, \\ Tel Aviv University, Tel Aviv 69978,
Israel\\ }

\date{\today}
\maketitle
\begin{abstract}
The adiabatic pumped current through an unbiased one dimensional
(1D) channel, connected to two 1D leads and subject to surface
acoustic waves (SAW), is calculated exactly for non-interacting
electrons. For a broad range of the parameters,
quantum interference generates a staircase structure of the
time-averaged current, similar to experimental observations. This
corresponds to integer values (in units of electronic charge) of
the charge pumped during each period of the SAW. We also find
staircases for higher harmonics. Quantum interference can thus
replace Coulomb blockade in explaining the pumped charge
quantization, particularly in the SAW experiments.

\end{abstract}
\pacs{73.23.-b,73.63.Rt,73.50.Rb,73.40.Ei}

\begin{multicols}{2}

Mesoscopic devices which exhibit quantized  dc currents may serve
as accurate current standards, to be used in single-electron
metrology.\cite{flensberg} Theoretically, such quantization was
first suggested by Thouless,\cite{Th83} for non-interacting
electrons subject to a slowly moving periodic potential which is
superimposed on an infinite periodic system. This model is an
example of {\it adiabatic quantum charge pumping},\cite{He91} a
phenomenon which attracted much recent theoretical
interest.\cite{AG99,PT,Br98,AA98,Z99,SAA00,S00,Ma00,LEW,Lev00,W00,Mo01,Sh01,Ma01,wei2}
In such pumping, an oscillating potential, with period
$\tau=2\pi/\omega=1/f$, generates a time-averaged current $\bar I$
{\it without any bias between the two terminals}. Under ideal
conditions, $\bar I$ has the quantized values ${\cal N}e f$, with
an integer ${\cal N}$, where $e$ is the electron charge. This
implies that the charge $Q$ transferred during time $\tau$ is
$Q={\cal N}e$. The direction of the current is determined by phase
shifts between the oscillating potential at different locations.

Quantized pumped currents have been observed in two types of
experiments. In the first, the periodically moving potential is
provided by the piezoelectric potential of {\it surface acoustic
waves} (SAW) generated in a quasi-one-dimensional (1D) GaAs-AlGaAs
channel, \cite{Sh96,Ta97} and the observed average acoustoelectric
current exhibits steps between plateaus at ${\cal N}e f$ as
function of either the gate voltage, $V$, or the amplitude of the
SAW, $P$. In the second, out-of-phase oscillatory voltages were
applied to the barriers connecting a quantum dot (QD) to the
leads, in {\it turnstile-type devices}.\cite{Sw99,Po92,Kou91} In
both cases, the apparent quantization of $Q$ was attributed to the
Coulomb blockade which dictates the integral  number of electrons
`carried' by the moving potential well. Although such explanations
may apply to almost closed dots, electron-electron interactions
should be less important in {\it almost open} dots (as used e. g.
in \onlinecite{Sw99}).

Adiabatic unbiased quantum pumping for {\it non-interacting
electrons} was recently studied for the turnstile-like
geometry.\cite{Lev00} When the Fermi energy in the leads $E_F$
aligns with the energy of the quasi-bound state in the QD, $Q/e$
was found to be close to one. Along similar lines, \cite{W00} the
existence of resonant states in a QD was found to greatly enhance
the magnitude of $Q$, causing it to change sharply as function of
some parameters. Here we present a simple model for the SAW
experiment, which exhibits steps and integer plateaus in $Q$, as
function of parameters, for {\it non-interacting electrons}. Each
step in $Q$ is accompanied by a narrow peak in the otherwise very
small time-averaged transmission $\bar {\cal T}$. The calculated
staircase plots (e. g. Fig. \ref{fig1Q} below), and the
accompanying peaks in $\bar {\cal T}$, are qualitatively
reminiscent of the quantum Hall transverse and longitudinal
resistances. Indeed, both effects follow from steps in Berry-like
phases of the quantum wave functions, resulting only from {\it
quantum interference}.\cite{Th83,AG99,PT} Interestingly, our
results are also similar to those observed in the SAW experiments.

Unlike the turnstile-like case, the piezoelectric potential
generated by the SAW oscillates with time {\it everywhere} inside
the nanostructure, ${\cal H}_{\rm SAW}({\bf r},t)=P \cos(\omega
t-{\bf q}\cdot{\bf r})$, with the SAW wavevector ${\bf q}$, while
being heavily screened by the 2DEG forming the leads. SAW usually
satisfy the adiabatic conditions, as $\hbar \omega$ is small
compared to the relevant electronic energy scales.\cite{Th83} In
the absence of bias, SAW generate a non-zero average current, in
the direction of ${\bf q}$. A realistic treatment of the
experimental geometry\cite{LEW} only allowed a calculation at low
$P$, yielding $Q \propto P^2$. The decay of ${\cal H}_{\rm SAW}$
in the wide banks of the channel is also difficult to treat
exactly. In order to obtain a simple solvable model, we simplify
all of these by a 1D tight-binding model. The `channel' is
 made of
sites $n=1,~2,~...,~N$. It is connected to two 1D `leads' via
sites $1$ and $N$. The leads, which connect to electron reservoirs
(with the same chemical potential), are described by ${\cal
H}_0=-J \sum_n (|n \rangle \langle n+1 |+ hc)$ for $n<0$ and
$n>N$, with  eigenfunctions like $e^{\pm ikan}$ and energies
$E=-2J \cos ka$, where $a$ is the lattice constant. The `channel'
is modeled by
\begin{eqnarray}
{\cal H}_{\rm osc}&=&\sum_n \{\epsilon_n (t)|n \rangle \langle
n|-J_n(|n \rangle \langle n+1 |+ hc) \}. \label{Hosc}
\end{eqnarray}
with $J_n=J_D$ inside the channel, i. e. for $1 \le n \le N-1$,
and $J_0=J_\ell$, $J_N=J_r$ for the `contacts' with the leads. To
model ${\cal H}_{\rm SAW}({\bf r},t)$, we choose the site energies
inside the channel ($1 \le n \le N$) as
\begin{eqnarray}
\epsilon_n(t)=V+ P \cos[\omega t- qa(n-n_0)], \label{eps}
\end{eqnarray}
where $V$ represents the gate voltage and $P>0$, so that
$\epsilon_n$ has a maximum (or minimum) in the center of the
channel $n_0=(N+1)/2$ at $t=0$ (or $\tau/2$). This potential acts
only inside the channel, imitating the screening of the SAW
outside.

Figure \ref{fig1Q} shows $Q$ (in units of $e$) versus $V$ for
$N=6$ and some `optimal' parameters, at zero temperature ($T=0$).
One clearly observes three plateaus at each end, with $Q/e$ very
close to $\pm {\cal N}$ and ${\cal N}=0,~1,~2,~3$. Generally, we
observe sharp plateaus up to ${\cal N}=N/2$, with possibly several
additional rounded peaks or spikes. The steps, at $V_{\cal N}$,
between these plateaus appear to be equidistant; for large $N$ we
show below that
\begin{eqnarray}
V_{\cal N}& \approx &E_F \pm\bigl (P+2 J_D-\Delta({\cal
N}+\frac{1}{2})\bigr ),
\nonumber\\
\Delta& = &qa \sqrt{2 P J_D}, \label{delta}
\end{eqnarray}
 i. e. the steps move outwards (left and right) and broaden with increasing
 $P$ and $J_D$, as also happens qualitatively in the
 experiments.\cite{Sh96,Ta97} It would be nice to examine this
 detailed quantitative prediction experimentally.
 As one expects, $Q=0$ when any one of $P,~J_L,~J_D,~qa,~ka$ or
$E_F-V$ vanishes. Indeed, the steps become rounded and decrease
gradually as $P,~E_F-V$ or $qa$ approach zero. The rounding begins
at the larger ${\cal N}$'s; the plateaus at ${\cal N}=\pm 1$
disappear last. The results remain robust for a wide range of
$ka,~J_L$ and $J_D$, provided $0<J_L^2/J \le J_D \ll P,~|E_F-V|$;
Fig. \ref{param} shows the effects of increasing $J_D$, and of
going to the limit of a completely open channel ($J_L=1$ and
$ka=\pi/2$). In the latter case, electron-electron interactions
should be unimportant! As Eq. (\ref{delta}) implies, we also
observe steps and plateaus for $Q$ versus $P$ at fixed $V$: at low
$P$, one starts with $Q \propto P^2$, but $Q$ remains very small
up to $P_0=E_F-V-2 J_D+\Delta/2$. Above $P_0$ one observes $N/2$
steps, at intervals $\Delta$ (which now increases with $P$), and
then a gradual decrease towards zero. Thus, both $V$ and $P$ can
be used as triggers for on-off switching of the pumped current.

The `optimal' graph in Fig. \ref{fig1Q} was derived at
$qa=\pi/[2(N-1)]$, corresponding to a total phase shift of $\pi/2$
between the end points ($n=1$ and $N$), or to a SAW wave length
$\lambda$ equal to $4L$, where $L=(N-1)a$ is the length of the
channel. Figure \ref{fqa} shows results for $\lambda/L=1,~1.5,~2$
and $8$; clearly, the staircase deteriorates as  $\lambda$ moves
away from $4L$, in both directions.
Even as the higher plateaus deteriorate, the plateaus at ${\cal
N}=\pm 1$ remain quite robust, down to $\lambda \sim L$ as
apparently used in the experiments.\cite{Sh96} Note that in the
latter, the SAW decays towards the ends of the channel, and
therefore the relevant $L$ may be smaller than estimated, bringing
the experiments closer to our `optimal' range.

Our calculations involve three main steps. In the first, we use
the adiabatic approximation (neglecting high orders in
time-derivatives of the wave functions) to calculate the unbiased
current. Such calculations usually employ the Brower
formula\cite{Br98}, which involves integrals over the
multidimensional space of the time-dependent potentials. Here we
use an equivalent formula, which is more convenient for a general
${\cal H}_{\rm osc}$.
  If $|\chi_\alpha^t \rangle$ denotes
the instantaneous scattering solution at time $t$, which results
from an incoming wave $|w_\alpha^-\rangle$ from lead
$\alpha~(=\ell,r)$ with energy $E$, then the instantaneous
current from left to right is\cite{long}
\begin{eqnarray}
I^t_\ell&=&\frac{e}{2\pi}\int dE \Bigl (-\frac{\partial
f}{\partial E} \Bigr )\langle \chi_\ell^t|\dot {\cal H}_{\rm
osc}|\chi_\ell^t\rangle , \label{current}
\end{eqnarray}
where $f(E)$ is the Fermi distribution (which is the same in both
leads, in the unbiased limit), and the dot denotes the time
derivative. Below we use $T=0$, i. e. $E=E_F$.
Using (\ref{current}), the charge pumped in one period becomes
$Q_\ell=\int_{-\tau/2}^{\tau/2} dt I^t_\ell$. For a symmetric
channel ($J_\ell=J_r$) one has $Q_\ell=-Q_r= (Q_\ell-Q_r)/2\equiv
Q \equiv \bar I \tau$.

 The second
step involves finding $|\chi_\alpha^t \rangle$. This is similar to
a static scattering solution: we write
\begin{eqnarray}
\langle n|
\chi_{\ell}^t\rangle&=&A_{0,\ell}e^{ikan}+A_{\ell}e^{-ikan},\ \ n
\le 0,
\nonumber\\
\langle n|\chi_{\ell}^t \rangle &=&B_{\ell}e^{ikan},\ \ n \ge N+1,
\label{chil}
\end{eqnarray}
with $A_{0,\ell}=1/\sqrt{2J \sin ka}$ (for a unit incoming flux).
The Schr\"{o}dinger equations at the sites $n=0$ and $n=N+1$ now
yield $A_{0,\ell}e^{ika}+A_{\ell}e^{-ika}=\phi_{\ell}(1)J_\ell/J$
and $B_{\ell}e^{ikNa}=\phi_{\ell}(N)J_r/J$, where $\{
\phi_{\ell}(n) \}$ are the amplitudes of the scattering solution
inside the nanostructure, which obey
\begin{eqnarray}
&&\Sigma_{n'} {\cal M}_{nn'}\phi_{\ell}(n')
=2\delta_{n,1}i\sin ka e^{ika}J_{\ell}A_{0,\ell},\nonumber\\
&&{\cal M}_{n,n'} \equiv \bigl (g^{-1}(E)\bigr )_{n,n'}=
E \delta_{n,n'}-\bigl ({\cal H}_{\rm osc} \bigr )_{n,n'}\nonumber\\
&&+\delta_{n,n'}e^{ika}\bigl
(\delta_{n,1}J_{\ell}^{2}+\delta_{n,N} J_{r}^{2}\bigr )/J,
\label{MM}
\end{eqnarray}
for $1 \le n,~n' \le N$.
The solution of these
equations is

\begin{eqnarray}
\phi_{\ell}(n)&=&e^{ika}A_{0,\ell}2i\sin ka J_{\ell}g_{n,1},
\nonumber\\
 A_{\ell}&=&e^{2ika}A_{0,\ell}[2i\sin ka
g_{1,1}J_\ell^2/J-1], \nonumber\\
B_{\ell}&=&e^{ika(1-N)}A_{0,\ell}2i\sin ka g_{N,1}J_\ell J_r/J,
\label{ABphi}
\end{eqnarray}
with similar expressions for $\phi_r(n),~A_r$ and $B_r$. The last
equation in (\ref{ABphi}) yields the instantaneous (`normal')
transmission, ${\cal T}^t=4|g_{N,1}|^2(J_\ell J_r/J)^2 \sin^2 ka$.
In the cases of interest here, ${\cal T}^t$ is usually very small.
However, ${\cal T}^t$ has local peaks (as function of $\cos \omega
t$) at the $N$ poles of $g_{N,1}$, i. e. the zeroes
of $D(\cos \omega t)=\det {\cal M}$. The time-average $\bar{\cal
T}=\int_{-\tau/2}^{\tau/2} dt{\cal T}^t/\tau$ thus exhibits peaks
wherever such poles occur within the period $\tau$.

 If
${\cal H}_{\rm osc}$ depends on time only via $\epsilon_n$, then
Eq. (\ref{current}) yields
\begin{equation}
Q_\ell=\frac{e J_\ell^2 \sin ka}{\pi J} \int_{-\tau/2}^{\tau/2} dt
\sum_{n=1}^N \dot{\epsilon}_{n}|g_{n,1}|^{2}.
\label{int}
\end{equation}
and $Q_\ell$ shows singularities whenever $\cos \omega t$ comes
close to one or more zeroes of $D$ within the integration. For
small $(J_\ell^2+J_r^2)/J$, these poles have small imaginary
parts,
and $Q$ exhibits large changes whenever $\cos \omega t$ passes
near such a pole. These steps occur exactly when $\bar {\cal T}$
has spikes, originating from the same poles.

 In the third step, we apply the above adiabatic equations to our
specific 1D channel model.  The $N \times N$ matrix ${\cal
M}=g^{-1}(E)$ is now tridiagonal, with ${\cal
M}_{n,n}=E-\epsilon_n+e^{ika}(\delta_{n,1}J_l^2+\delta_{n,N}J_r^2)/J$
and ${\cal M}_{n,n \pm 1}=J_D$. Since $\dot \epsilon_n=-\omega P
\sin[\omega t-qa(n-n_0)]$, $I_\ell^t$ is equal to $\omega$ times a
function of $\omega t$, $\bar I \propto \omega$ and $Q$ is
independent of $\omega$.
For our symmetric channel ($J_\ell=J_r$), $D$ is an order-$N$
polynomial in $\cos \omega t$; we calculate $Q$ by rewriting the
integrand as a sum over its $2N$ complex poles in $\cos \omega t$,
and using analytic expressions for $\int_0^\pi du u^m/(\cos u
-z)$.

It is interesting to follow these poles as function of the various
parameters. For this purpose, we show in Fig. \ref{QQt} the
partial charge $Q_\ell(t)$, resulting from integration of Eq.
(\ref{int}) only up to $t<\tau/2$, at different values of $V$. As
$V$ increases through $V_1$, $Q_\ell(t)$ suddenly exhibits a step
from zero to one, which appears at $t=0$. This step corresponds to
a pole in $I^t_\ell$, which enters near $\cos \omega t=1$. As $V$
increases, this step moves to the left. At $V=V_2$, another step
(from 1 to 2) enters at $t=0$. For large $N$, the time interval
between two consecutive steps is roughly equal to
$\delta=qa/\omega$. As $V$ increases further, both steps move to
the left, and at $V=V_3$ a step from 2 to 3 enters at $t=0$. After
a narrow intermediate complex state, there begin to enter steps of
$-1$, until at $V=E_F$ we observe exactly $N/2$ steps of $+1$
followed by $N/2$ steps of $-1$, ending up with $Q=0$. A similar
build-up of (negative) steps occurs when one starts at large
positive $V$, and follows $V$ down through $V_{-{\cal N}}$, except
for the fact that now the new steps show up at $t=\tau/2$, i. e.
$\cos \omega t=-1$. The physical interpretation of these results
is clear: unlike the Coulomb blockade picture, where ${\cal N}$
electrons move together from left to right, carried by a single
minimum of the moving potential, in the SAW case treated here
$Q_\ell$ changes by steps of 1, implying {\it separate motion of
individual electrons}, building up to ${\cal N}$ after a full
period. The picture is particularly interesting near $V=E_F$:
during a period, $Q_\ell(t)$ exhibits several positive steps, and
then an equal number of negative steps, ending with no net pumped
charge.

The step-like time dependence of $Q_\ell(t)$ immediately implies
the appearance of {\it higher  harmonics} (in $\omega$) of the
pumped current: for $V$ in the ${\cal N}$'th plateau, we
approximate $I^t_\ell \approx e \sum_{{j}=1}^{\cal N}
\delta(t-t_{j})$, with $t_{j}=t_0(V)+(j-1)\delta$. The pumped
current then becomes $I_\ell^t=ef \{{\cal N}+2 \sum_{m=1}^\infty
\cos[m \omega(t-t_0-({\cal N}-1)\delta/2)]{\cal I}(m,{\cal N})\}$,
with
\begin{equation}
{\cal I}(m,{\cal N}) \approx \sin({\cal N}m \omega
\delta/2)/\sin(m \omega \delta/2).
\end{equation}
Thus, this $m$'th harmonic amplitude also exhibits a staircase
structure. Although $|{\cal I}(m,1)|^2 \equiv 1$, the plateaus at
higher ${\cal N}$ oscillate with $m$. It is interesting to note
that harmonics with $m \ne 0$ survive in the intermediate range $V
\sim E_F$: as seen from the center Fig. \ref{QQt}, one then has a
difference like $\cos[m \omega(t-t_1)]-\cos[m \omega(t-t_2)]$. It
would be very interesting to study ${\cal I}(m,{\cal N})$
experimentally.

Since the steps in $Q(V)$ arise when poles appear (or disappear)
when $\cos \omega t=\pm 1$, we confirmed that the $V_{\cal N}$'s
are the solutions of $D(V, \cos \omega t=\pm 1)=0$. Taking $t=
\tau/2$, this equation is equivalent to the set of equations
\begin{eqnarray}
&&\bigl (E-V +P \cos[(n-n_0)qa]\bigr )\phi(n)=
\nonumber\\
&&-J_D[\phi(n+1)+\phi(n-1)] \label{mathieu}
\end{eqnarray}
 for $1<n<N$, with modified
equations for $n=1$ and $n=N$. Except for the boundaries, these
are Mathieu's equations, with the symmetric potential having a
minimum at $n=n_0$. This can be written as
\begin{eqnarray}
&&-J_D[\phi(n+1)+\phi(n-1)-2\phi(n)]\nonumber\\
&&+P(1-\cos[(n-n_0)qa])\phi(n)={\textsf{{E}}}\phi(n), \label{osc}
\end{eqnarray}
with $\textsf{{E}}=E-V+P+2 J_D$. For our `optimal'
$qa=\pi/[2(N-1)]$, one has $|n-n_0|qa \le \pi/4$, so that the
replacement $1-\cos[(n-n_0)qa] \approx \frac{1}{2}(n-n_0)^2(qa)^2$
forms an excellent approximation. The low lying eigenenergies of
the isolated channel then correspond to localized states around
$n_0$, which are not very affected by the boundaries. For large
$N$, we also use $[\phi(n+1)+\phi(n-1)-2\phi(n)] \approx
\partial^2 \phi/\partial n^2$. Eq. (\ref{osc}) then
becomes the Schr\"{o}dinger equation for the harmonic oscillator,
yielding the eigenvalues ${\textsf{{E}}}_{\cal N}=\Delta({\cal
N}+1/2)$, with integer ${\cal N}$, i. e. Eq. (\ref{delta}) with
the upper sign. The solution for the other sign, associated with
poles entering at $t=0$, follows the replacement $\phi(n)
\rightarrow (-1)^n\phi(n)$. These approximate values agree very
well with our direct solutions of $D(V,\pm 1)=0$ and with the
steps in $Q$, even for relatively small $N$ (e. g. Fig.
\ref{fig1Q}).

A few more comments are in place: (a) Although our calculated
plateaus look like constant integers, in practice the function
$Q(V)$ remains slightly below ${\cal N}$, reaching a smooth
maximum around the middle of the `plateau'. For $N=6$, the
difference $({\cal N}-Q(V))$ is of order .0001. This difference
becomes smaller for larger $N$.   (b) Our robust plateaus at
integer values occur only for the model presented here, where the
amplitude $P$ has exactly the same value for all $n$ in the
channel. Modulation of $P$ in space, e. g. due to a screening
decay, due to a reflected SAW,\cite{Ta97} or due to random
energies $\{V_n\}$, gives plateaus at non-integer values or round
the steps. This, and a treatment of more complex nanostructures,
will be reported elsewhere.\cite{long} (c) Our formalism allows
for $T>0$, where we only expect some rounding of the steps. (d)
Eq. (\ref{delta}) for $\Delta$ remains true in the limit $N
\rightarrow \infty,~a \rightarrow 0,~(N-1)a=L$, when $J_D a^2
\rightarrow \hbar^2/(2 m^\ast)$.  (e) Finally, we emphasize again
that all of our results are valid only in the adiabatic limit.


 We thank Y. Imry, Y. Levinson and P. W\"{o}lfle for helpful
conversations. This research was carried out in a center of
excellence supported by the Israel Science Foundation, and was
supported in part by the National Science Foundation under Grant
No. PHY99-07949 and by the Einstein Center at the Weizmann
Institute.

\begin{figure}
\leftline{\epsfclipon\epsfxsize=3.1in\epsfysize=2in\epsffile{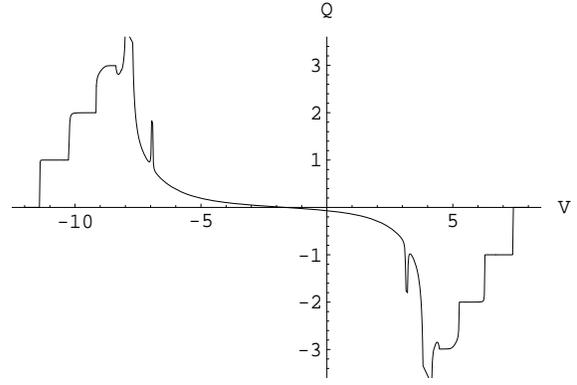}}
\vskip .4truecm \caption{Pumped charge $Q$ (in units of $e$)
versus the gate voltage $V$ (in units of $J$) for $N=6,~P=8J$,
$J_D=J,~J_\ell=J_r=J_L=.4J,~qa=\pi/10,~ka=\pi/100$.} \label{fig1Q}
\end{figure}

\begin{figure}
\leftline{\epsfclipon\epsfxsize=3.1in\epsfysize=1.2in\epsffile{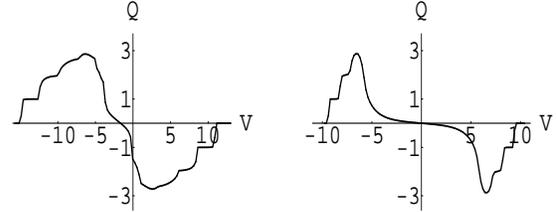}}
\vskip .4truecm \caption{Same as Fig. \ref{fig1Q}, but with
$J_D=3J$ (left) and $J_L=J,~ka=\pi/2$ (right).} \label{param}
\end{figure}

\begin{figure}
\leftline{\epsfclipon\epsfxsize=3.1in\epsfysize=2in\epsffile{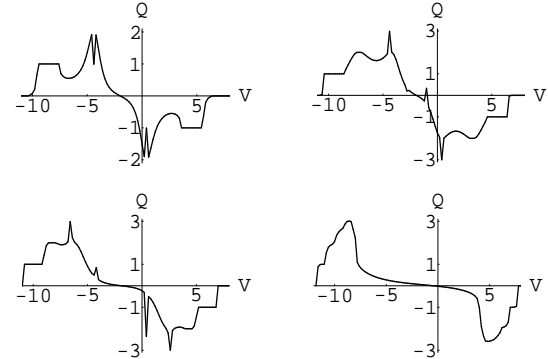}}
\vskip .5truecm \caption{Same as Fig. \ref{fig1Q}, but with
$qa=2\pi/5,4\pi/15,\pi/5$ and $\pi/20$, corresponding to
$\lambda/L=1,1.5,2$ and $8$.} \label{fqa}
\end{figure}

\begin{figure}
\leftline{\epsfclipon\epsfxsize=3.1in\epsfysize=.6in\epsffile{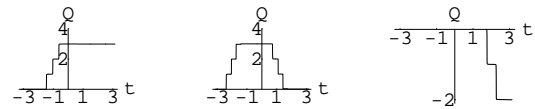}}
\vskip .4truecm \caption{Partial pumped charges $Q_\ell$, in units
of $e$, up to time $t$ within a period, for the same parameters as
in Fig. \ref{fig1Q}, with $V/J=-8.6,~-2$ and $5.3$. The $t$-axis
shows $\omega t$ between $-\pi$ and $\pi$.} \label{QQt}
\end{figure}

\end{multicols}

\end{document}